\documentclass[3p,onecolumn]{elsarticle}

\usepackage{graphicx}
\usepackage{dcolumn}
\usepackage{pifont}
\usepackage{bm}
\usepackage{multirow}
\usepackage{amsmath}
\usepackage{float}
\usepackage{txfonts}
\usepackage[version=3]{mhchem}

\def\mathbi#1{\textbf{\em #1}}

\journal{Journal of Power Sources}

\begin{document}
\title{Revealing the stability and efficiency enhancement in mixed halide perovskites \ce{MAPb(I_{1-x}Cl_x)3} with {\it ab initio} calculations}

\author[kimuniv-m,kimuniv-n]{Un-Gi Jong}
\author[kimuniv-m]{Chol-Jun Yu\corref{cor}}
\ead{ryongnam14@yahoo.com}
\author[kimuniv-n]{Yong-Man Jang}
\author[kimuniv-m]{Gum-Chol Ri}
\author[kimuniv-m]{Song-Nam Hong}
\author[kimchaek]{Yong-Hyon Pae}

\cortext[cor]{Corresponding author}

\address[kimuniv-m]{Department of Computational Materials Design, Faculty of Materials Science, Kim Il Sung University, \\ Ryongnam-Dong, Taesong District, Pyongyang, Democratic People's Republic of Korea}
\address[kimuniv-n]{Natural Science Centre, Kim Il Sung University, Ryongnam-Dong, Taesong District, Pyongyang, Democratic People's Republic of Korea}
\address[kimchaek]{Institute of Physics Engineering, Kim Chaek University of Technology, Central District, Pyongyang, Democratic People's Republic of Korea}

\begin{abstract}
A little addition of Cl to \ce{MAPbI3} has been reported to improve the material stability as well as light harvesting and carrier conducting properties of organometal trihalide perovskites, the key component of perovskite solar cell (PSC). However, the mechanism of performance enhancement of PSC by Cl addition is still unclear. Here, we apply the efficient virtual crystal approximation method to revealing the effects of Cl addition on the structural, electronic, optical properties and material stability of \ce{MAPb(I_{1-x}Cl_x)3}. Our {\it ab initio} calculations present that as the increase of Cl content cubic lattice constants and static dielectric constants decrease linearly, while band gaps and exciton binding energies increase quadratically. Moreover, we find the minimum of exciton binding energy at the Cl content of 7\%, at which the chemical decomposition reaction changes coincidentally to be from exothermic to endothermic. Interactions among constituents of compound and electronic charge transferring during formation are carefully discussed. This reveals new prospects for understanding and designing of stable, high efficiency PSCs.
\end{abstract}


\maketitle

\section{\label{sec:intro}Introduction}
The world has been searching for inexpensive, renewable and clean energy sources with growing concerns on more and more energy demands, global warming and serious climate change. Among various options for such energy sources, solar power should have undoubtedly the greatest potential due to its most abundance and eternal nature, despite the drawbacks such as being variable in time and widespread in space, which require to create new technologies for effective solar energy harvesting. Recently, photovoltaic (PV) technology, known to be a promising route for harnessing solar power, has attained the new prosperity since a breakthrough occurred by an invention of perovskite solar cell (PSC) using organometal trihalide perovskite (OTP) as a light harvester~\cite{Kojima,Liu13}. A great attraction of PSCs comes from remarkably low energy cost compared with other conventional solar cells. The reduction of energy cost in PSCs may be associated with low cost of substrate materials, easy manufacturing process, and high performance of device given by efficiency and stability~\cite{Song,Xiao}, which promise for PV energy to become competitive with fossil fuels~\cite{Luo,Sum}. This is due to the unique optoelectronic properties of OTPs, typically methylammonium lead trihalide perovskite \ce{MAPbX3} (MA=\ce{CH3NH3}, X=Cl, Br, I) that is the key component of PSC, such as large light absorption coefficients~\cite{Lindblad}, optimal band gaps~\cite{Frost14}, large and balanced carrier mobilities~\cite{Edri}, large carrier diffusion length~\cite{Xing,Stranks13}, and small exciton binding energies~\cite{Hakamata,Stranks15}.

However, there are still major challenges for OTPs such as poor material stability~\cite{Wang,NiuRev,Berhe} and toxicity of the lead~\cite{Noel}, which hinder practical use of PSCs. In particular, the issues of degradation of OTPs and stability of PSCs should be urgently resolved to achieve a long lifetime of PSCs with high conversion efficiency. Although PSCs consist of multilayers including light harvesting materials, electron and hole transport layers, counter electrodes, etc., the key problem of degradation is that \ce{MAPbI3}(s), the most popular OTP, may be readily decomposed into \ce{PbI2}(s) and \ce{MAI}(aq) or further \ce{CH3NH2}(aq) and \ce{HI}(aq) under different ambient conditions during operation, such as moisture~\cite{Han,Christians} or oxygen in air~\cite{Zhou14}, ultraviolet light~\cite{Leijtens,SongUV}, and thermal effect~\cite{Bi,Dualeh}. In spite of the very importance and urgency, the mechanism of degradation by each of external actions or synergy action is not yet fully understandable.

In order to resolve poor material stability of \ce{MAPbI3}, any modification in the position of component ions could be thought~\cite{Berhe,NiuRev}. One successful technique is to replace organic MA cation with larger size organic cation like formamidinium ion (\ce{FA+}=\ce{HC(NH2)2+}), which was found to increase the material stability~\cite{AharonFA,Eperon}. When replacing Pb with Sn, however, the stability reduces significantly due to a possible oxidation of \ce{Sn^{2+}}, although some merits like non-toxicity are expected~\cite{Noel}. Then, the most widely studied substitution could be in halogen ion position; the stability of \ce{MAPbI3} can be remarkably improved when mixing with a small fraction of Br or Cl~\cite{Noh,Lee,Dimesso}. In the case of \ce{MAPb(I_{1-x}Br_x)3}, a high resistivity to humidity was found at $x\geq0.2$ from experiments~\cite{Noh,Aharon} and density functional theory (DFT) calculations~\cite{Mosconi,Jong,yucj16}. This might be associated with their compact and stable structures, since the replacement of larger I ions (2.20 \AA) with smaller Br ions (1.96 \AA) leads to a reduction in the lattice constant and a phase transition from the tetragonal to the cubic~\cite{Aharon}. A similar effect was observed when mixing with much smaller Cl ions (1.81 \AA)~\cite{Liu13,Liu16,Tripathi,Du,Chueh,Williams}. The resulting compact structure of \ce{MAPb(I_{1-x}Cl_x)3} could increase the resistivity of the material to moisture, light and temperature thanks to its higher binding compared to \ce{MAPbI3}. Nevertheless, there is no clear evidence of the role of chlorine insertion to enhance stability. In addition, these modifications using appropriate substitutions could improve other material properties of the perovskite related with efficiency, such as band gaps, light absorption coefficients and carrier mobilities. The insertion of Cl into \ce{MAPbI3}, for instance, improves the uniformity of its layer and results in an increase of recombination lifetime of carriers and electron diffusion length from 100 nm to 1 $\mu\text{m}$~\cite{Stranks13,Xing}. Although mixing halide in OTPs seems to have beneficial effects in terms of performance, the questions of what is the optimal Cl content in \ce{MAPb(I_{1-x}Cl_x)3} and why is still in debate due to a lack of systematic investigation.

In this work, we report an {\it ab initio} study on the mixed iodide-chloride perovskites \ce{MAPb(I_{1-x}Cl_x)3} using the virtual crystal approximation (VCA)~\cite{yucj07,Jong} method within DFT. As increasing the Cl content $x$, we first optimize the lattice parameters of cubic phases, and then, using the determined structures, calculate optoelectronic properties such as band gaps, effective masses of charge carriers, exciton binding energies and light absorption coefficients, which are used to demonstrate the advantages of charge carrier generation and transportation in these materials. At the end, we attempt to give a reason for the stability enhancement by mixing chlorine as well as a precise value of Cl content to enhance performance with a careful analysis of the calculated formation enthalpies and charge density differences.

\section{Computational methods}\label{method}
The {\it ab initio} calculations were carried out by applying the pseudopotential plane wave package ABINIT (version 7.10.2)~\cite{abinit09} using generalized gradient approximation (GGA) parameterized by Perdew-Burke-Ernzerhof (PBE)~\cite{pbe} with the semi-empirical van der Waals correction (vdW-D2)~\cite{Grimme}. The optimized norm-conserving, designed nonlocal pseudopotentials were generated with the OPIUM package~\cite{Rappe,Ramer}, using the following valence electronic configurations of atoms: H--1s$^1$, C--2s$^2$2p$^2$, N--2s$^2$2p$^3$, Cl--3s$^2$3p$^5$, I--5s$^2$5p$^5$, and Pb--5d$^{10}$6s$^2$6p$^2$. For the virtual atoms \ce{I_{1-x}Cl_x} from x=0 to 1 with an interval of 0.1, we applied the Yu-Emmerich extended averaging approach (YE$^2$A$^2$)~\cite{yucj07} to constructing their pseudopotentials, as it was proved for this method to give reasonable results for the mixed iodide-bromide perovskites~\cite{Jong}. A plane wave cutoff energy of 40 Ha and a 4$\times$4$\times$4 $k$-point grid were used, confirming the total energy convergence to be 5 meV per cell. When calculating the optoelectronic properties, a denser $k$-point mesh of 10$\times$10$\times$10 was used. We determined the optimal lattice parameters by calculating the total energies versus volumes of the unit cells, allowing the atomic relaxation with a force tolerance of 0.01 eV \AA$^{-1}$, and then fitting the $E$-$V$ data into the Birch-Murnaghan equation of state (EOS).

It was found from experiments that at room temperature, bulk \ce{MAPbI3} forms a tetragonal structure (space group $I4/mcm$) (and transforms into a cubic phase ($Pm$\={3}$m$) above $\sim56^\circ$C), whereas \ce{MAPbCl3} forms the cubic structure ($Pm$\={3}$m$) (and transforms into the tetragonal phase ($P4/mmm$) below $-96^\circ$C)~\cite{Poglitsch,Baikie}. After the incorporation of \ce{Cl-} into \ce{MAPbI3}, the cubic phase \ce{MAPb(I_{1-x}Cl_x)3} always occurred in meso-porous substrate~\cite{Colella,Williams}, although the tetragonal phase was occasionally found on compact TiO$_2$ substrate~\cite{Docampo}. It is worth noting that there is no critical difference between the tetragonal and cubic phases, except a slight rotation of \ce{PbI6} octahedra along the c-axis, and thus, the tetragonal phase can be treated as a pseudo-cubic phase ($Pm$) with $a^\star=a/\sqrt2, c^\star=c/2$~\cite{Eperon}. Based on these arguments, we treated the pseudo-cubic phase for all solid solutions \ce{MAPb(I_{1-x}Cl_x)3} (x=[0, 1]) in this work.

The macroscopic dielectric functions including the effect of electron-hole interaction, $\varepsilon_M(\omega)=\varepsilon_1(\omega)+i\varepsilon_2(\omega)$, were calculated within the Bethe-Salpeter approach~\cite{Onida} as implemented in the ABINIT code. Those were used to determine the photo-absorption coefficient, using the following equation:
\begin{equation}
\label{absorption}
\alpha(\omega)=\frac{2\omega}{c}\sqrt{\frac{[\varepsilon^2_1(\omega)+\varepsilon^2_2(\omega)]^{1/2}-\varepsilon_1(\omega)}{2}},
\end{equation}
where $c$ is the speed of light in vacuum. The static dielectric constants ($\varepsilon_s$) extracted at zero-frequency limit, together with effective masses of electron ($m_e^*$) and hole ($m_h^*$), were used to estimate the exciton binding energy within the weak Mott-Wannier model, using the following equation~\cite{Even14,Jong}:
\begin{equation}
 \label{eig_exciton}
E_b^\text{ex}=\frac{m_ee^4}{2(4\pi\varepsilon_0)^2\hbar^2}\frac{m_r^*}{m_e}\frac{1}{\varepsilon_s^2}\approx13.56\frac{m_r^*}{m_e}\frac{1}{\varepsilon_s^2}~(\text{eV}),
\end{equation}
where $1/m_r^*=1/m_e^*+1/m_h^*$ for effective reduced mass $m_r^*$. This model is validated when the extending radius $a_0^*=\varepsilon\frac{m_e}{m_r^*}a_0$ is larger than the lattice constant.

\section{Results and discussion}\label{result}
\subsection{Structure related properties}\label{struct}
\begin{figure*}[!th]
\begin{center}
\begin{tabular}{l@{\hspace{60pt}}c}
\includegraphics[clip=true,scale=0.185]{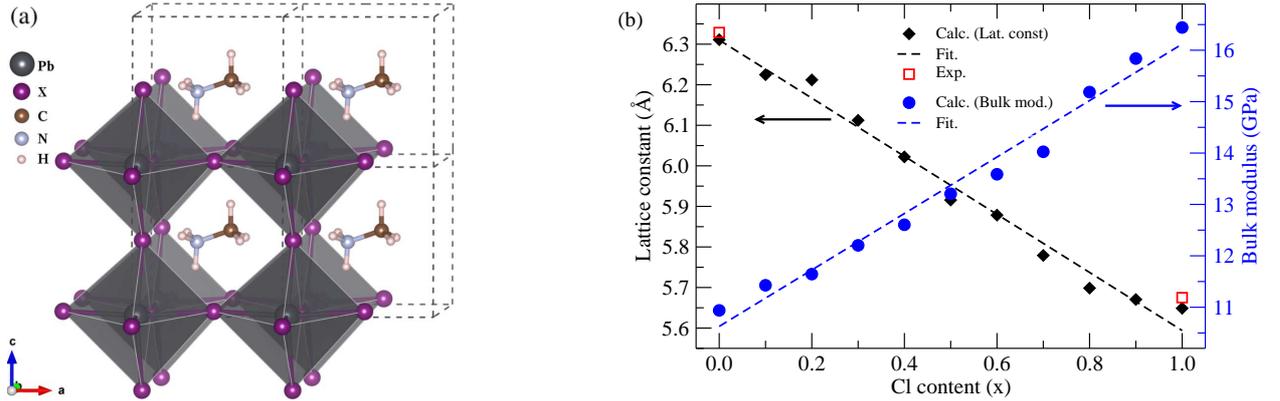} &
\includegraphics[clip=true,scale=0.47]{fig1b.eps}
\end{tabular}
\end{center}
\caption{\label{fig_model}(a) The crystalline structure of pseudo-cubic \ce{MAPbX3} (space group $Pm$) with [101] orientation of MA cation. (b) Lattice constants and bulk moduli as linear functions of Cl content x in \ce{MAPb(I_{1-x}Cl_x)3}. Dashed lines denote the fitting lines and Exp. data is from ref.~\cite{Poglitsch}. }
\end{figure*}

It was confirmed that there exist vdW interactions between the organic molecule and the inorganic matrix in the OTPs and they play a certain role in the description of material properties within the DFT framework~\cite{Motta,Geng,yucj16}. Accordingly, we first checked the validity of vdW inclusion by comparing the computed lattice constants from the PBE functional without vdW correction, with semi-empirical vdW correction (vdW-D2) and vdW functional (vdW-optPBE) to the experimental data of \ce{MAPbI3} and \ce{MAPbCl3}. As mentioned above, \ce{MAPbX3} (X=\ce{I_{1-x}Cl_x}, x=[0, 1]) crystals were modeled with a pseudo-cubic phase with [101] orientation of MA cation, as identified from experiments~\cite{Poglitsch,Baikie,Eperon} and verified to be the lowest energy configuration by our previous work~\cite{yucj16,Jong} (Fig.~\ref{fig_model}(a)).

We calculated the total energies of unit cells at fixed volume which varies from $0.9V_0$ to $1.1V_0$ with 10 intervals, while allowing full atomic relaxation, and then performed fittings of the $E-V$ data into the Birch-Murnaghan EOS, determining the equilibrium lattice constant and bulk modulus. To roughly estimate the structural stability, we analyzed the averaged Pb-X bond length, and the cohesive energy calculated as the difference between the total energy of bulk \ce{MAPbX3} and the sum of isolated constituent atoms' like $E_c=E(\ce{MAPbX3})-\sum_iE(i)$. The calculated equilibrium lattice constants, bulk moduli, averaged Pb-X bond lengths, and cohesive energies for \ce{MAPbI3}, \ce{MAPbCl3} and their solid solution \ce{MAPb(I_{0.5}Cl_{0.5})3} are summarized in Table~\ref{tab_strct}. It was found that, while the PBE overestimates the lattice constant as usual for other inorganic semiconductors, the vdW-D2 outputs the closest values compared to experiment and optPBE in between. Based on this check, the PBE+vdW-D2 method was used for further calculations. We see that, whatever the type of XC functional is, going from I to Cl, the lattice constant and Pb-X bond length decrease, while the bulk modulus and the magnitude of cohesive energy increase. However, there is no significant change in C-N bond length, indicating that the change of lattice does not affect the intramolecular behaviour of MA molecule but the Pb-X and \ce{PbX6}-MA bonding.
\begin{table}[!th]
\centering
\footnotesize
\caption{\label{tab_strct}Lattice constant $a$ (\AA), bulk modulus $B$ (GPa), averaged Pb-X bond length $d_\text{Pb-X}$ (\AA) and cohesive energy $E_c$ (eV)}
\begin{tabular}{l@{}ccccc}
\hline
 Material   & Property    & PBE & optPBE & vdW-D2 & Exp.~\cite{Poglitsch} \\
\hline
\ce{MAPbI3} & $a$   & 6.358 & 6.354 & \textbf{6.311} & 6.328 \\
            & $B$   & 14.61 & 13.53 & 10.94 &  \\
            & $d_\text{Pb-I}$  & 3.266 & 3.266 & 3.200 &  \\
            & $E_c$ & -3.72 & -3.37 & -3.58 &  \\
\hline
\ce{MAPb(I_{0.5}Cl_{0.5})3}& $a$    & 6.078 & 6.113 & 5.915 &  \\
            & $B$   & 17.66 & 18.21 & 13.21 &  \\
            & $d_\text{Pb-X}$ & 3.071 & 3.100 & 2.991 &  \\
            & $E_c$ & -3.78 & -3.44 & -3.65 &  \\
\hline
\ce{MAPbCl3}& $a$   & 5.868 & 5.801 & \textbf{5.649} & 5.675 \\
            & $B$   & 19.35 & 22.56 & 16.44 &  \\
            & $d_\text{Pb-Cl}$  & 3.044 & 2.934 & 2.860 &  \\
            & $E_c$ & -3.97 & -3.66 & -3.85 &  \\
\hline
\end{tabular} \\
\normalsize
\end{table}

By repeating the above procedure for pseudo-cubic \ce{MAPb(I_{1-x}Cl_x)3} as increasing the Cl content x, we obtained the systematic change of lattice constants and bulk moduli, as shown in Fig.~\ref{fig_model}(b) (EOS curves in Fig.S1\dag). They are of almost linear functions of Cl content without any anomaly in the middle, $a(x)=6.310-0.716x$ (\AA) for lattice constants and $B(x)=10.627+5.492x$ (GPa) for bulk moduli, showing a satisfaction of the Vegard's law for lattice constants. This indicates that mixing chloride with iodide perovskite induces a shrinking of lattice and an increase of bulk modulus, resulting in a strengthening of Pb-X bond and thus chemical stability.

\subsection{Electronic properties}\label{electro}
\begin{figure*}[!th]
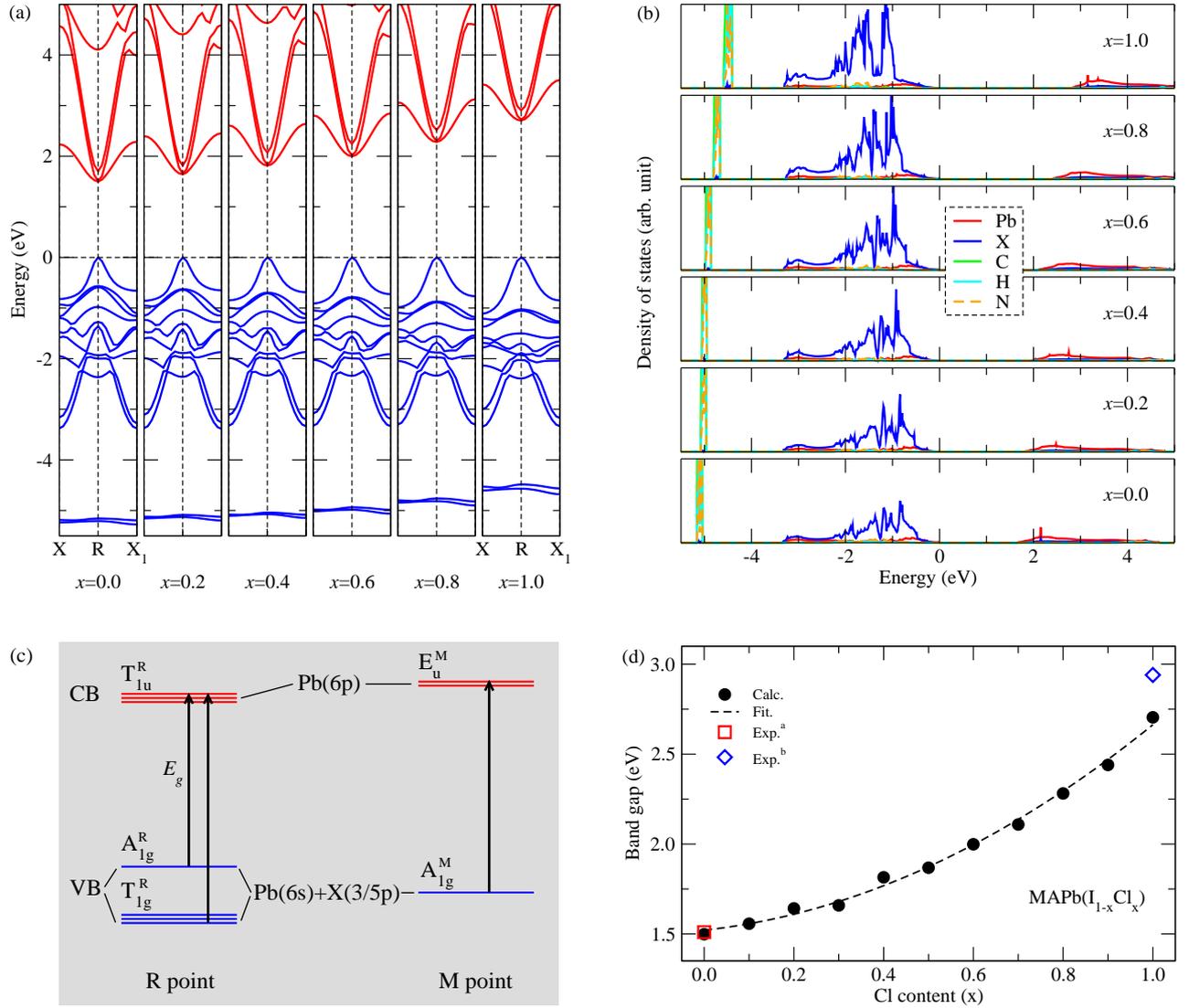

\begin{center}
\begin{tabular}{l@{\hspace{20pt}}r}
\includegraphics[clip=true,scale=0.47]{fig2a.eps} &
\includegraphics[clip=true,scale=0.47]{fig2b.eps}\\ \vspace{5pt} \\
\includegraphics[clip=true,scale=0.55]{fig2c.eps} &
\includegraphics[clip=true,scale=0.47]{fig2d.eps}
\end{tabular}
\end{center}
\caption{\label{fig_gap}Systematic change in electronic properties of \ce{MAPb(I_{1-x}Cl_x)3} (x=[0, 1]). (a) Electronic band structures around R point and (b) atomic resolved density of states, setting the valence band maximum to be zero. (c) Schematic energy level diagram depicted at R and M points with irreducible representations of the group and atomic contributions to the levels, and (d) band gaps with quadratic fitting line (dashed line). Exp$^a$ and Exp$^b$ are from ref.~\cite{Kim} and ref~\cite{Dimesso}, respectively.}
\end{figure*}

The electronic structures are crucial factors for solar cell application, since the light harvesting capability of the key materials, which mainly governs the performance of the solar cell, can be determined just from those. Accordingly, we investigated the electronic band structures and density of states (DOS) systematically as the increase of Cl content. As can be seen in Fig.~\ref{fig_gap}(a) and Fig.S2\dag, the band structures of the mixed perovskites \ce{MAPb(I_{1-x}Cl_x)3} from x=0.0 to x=1.0 are quite similar in overall with gradual increase of band gaps, which is in direct mode at the R point in the first Brillouin zone (1BZ). It is worth noting that the next direct band gap can occur at the M point in 1BZ (Fig.S2\dag). The symmetry analysis of the electronic states at the R and M points with group theory implies that the optically allowed electron transitions are at the R point between \ce{A_{1g}} valence band (VB) and 3-fold degenerated \ce{T_{1u}} conduction band (CB) or \ce{T_{1g}} VB and \ce{T_{1u}} CB, and at the M point between \ce{A_{1g}} VB and 2-fold degenerated \ce{E_{u}} CB, as depicted in Fig.~\ref{fig_gap}(c)~\cite{Even14}.

Atomic resolved and partial DOSs are shown in Fig.~\ref{fig_gap}(b) and Fig.S3\dag. The top of the valence band at the R and M points is composed of strong antibonding coupling of Pb $6s$ with $X$ $3/5p$ orbitals, making upper VBs dispersive. On the other hand, the bottom of the conduction band at the R and M points is predominantly Pb $6p$ orbitals with negligible antibonding coupling with $X$ $3/5s$ orbitals, indicating the ionicity of \ce{MAPbX3}. In addition, the occupied molecular orbitals of MA cation, found deep $\sim$5 eV below the VB maximum, have no significant coupling with other orbitals, which reminds the aforementioned weak vdW interaction between organic MA cation and the inorganic PbX$_6$ matrix. When increasing the Cl content in the mixed \ce{MAPbX3}, moreover, the molecular orbitals of MA cation get closer to the hybridized orbitals of PbX$_6$ matrix due to getting stronger MA--PbX$_6$ interaction~\cite{Motta}, which may cause an upward shift of the CB orbitals, resulting in the increase of band gaps from \ce{MAPbI3} to \ce{MAPbCl3}.

From the band structure, the direct band gap of pseudo-cubic \ce{MAPbI3} is estimated to be 1.50 eV at the R point, which is in good agreement with the experimental data~\cite{Kim}. It is well known that this agreement is accidental coincidence thanks to the compensation of GGA underestimation by the lack of SOC~\cite{Even,Motta,Giorgi,Even14}. In fact, the large SOC of lead atom can induce a split of the CB into a 2-fold degenerated \ce{E_{1/2g}} and a 4-fold degenerated \ce{F_{3/2u}} state at the R point, making the band gap energy reduced to be like 0.5 eV~\cite{Even14}. Although this severe underestimation by the SOC effect can be cured by linking with GW method in defiance of heavy computational cost~\cite{Umari}, we could be satisfied with the results from the PBE+vdW-D2 without SOC. When replacing I atom with lighter Cl atom, {\it i.e.}, for \ce{MAPbCl3}, the SOC effect seems to become weaker, leading to a slight underestimation of the band gap (2.70 eV) compared to the experimental value (2.94 eV)~\cite{Dimesso} (Table~\ref{tab_gap}).
\begin{table}[!th]
\small
\centering
\caption{\label{tab_gap}Band gap (eV) and effective masses of electron and hole}
\begin{tabular}{l@{}cc@{}ccccc}
\hline
    & \multicolumn{2}{c}{Band gap} && \multicolumn{4}{c}{Effective mass} \\
\cline{2-3}  \cline{5-8}
Material &  Calc.  & Exp. & & $m_e^*$ & $m_e^{*c}$ & $m_h^*$ & $m_h^{*c}$ \\
\hline
\ce{MAPbI3} & 1.50 & 1.50$^a$ && 0.19 & 0.23 & 0.24 & 0.29 \\
\ce{MAPb(I_{0.5}Cl_{0.5})3}& 1.87 &   && 0.27 &   & 0.34 & \\
\ce{MAPbCl3}& 2.70 & 2.94$^b$ && 0.34 &  & 0.47 & \\
\hline
\end{tabular} \\
$^a$~Ref.~\cite{Kim}, $^b$~ref.~\cite{Dimesso}, $^c$~PBE+SOC calculation~\cite{Giorgi}
\normalsize
\end{table}

With the gradual increase of the Cl content in the mixed perovskites \ce{MAPbX3}, the band gaps are likely to increase quadratically as shown in Fig.~\ref{fig_gap}(d). The data can be interpolated into the following quadratic fuction~\cite{Jong}:
\begin{equation}
E_g(x)=E_g(0)+[E_g(1)-E_g(0)-b]x+bx^2
\end{equation}
with $E_g(0)=1.521$ eV, $E_g(1)=2.663$ eV, and $b=0.873$ eV for the bowing parameter. Since the bowing parameter reflects the fluctuation degree in the crystal field and the nonlinear effect from the anisotropic binding~\cite{Noh}, its larger value compared to 0.33 eV in \ce{MAPb(I_{1-x}Br_x)3}~\cite{Jong} indicates larger compositional disorder and low miscibility between \ce{MAPbI3} and \ce{MAPbCl3}. The replacement of some I ions by Cl ions with smaller ionic radius leads to an enhancement of interaction between Pb and X within PbX$_6$ octahedra, resulting in the contraction of lattice as well as the decrease of Pb-X bond length. This causes the increase of band gaps and thus a spoil of the light harvesting properties on the increment of Cl content.

Within the parabolic approximation, we calculated the effective masses of electron ($m^*_e$) around the top of VB and hole ($m^*_h$) around the bottom of CB at the R point by numerically processing the band structures. As listed in Table~\ref{tab_gap} and shown in Fig.S4\dag, they increase linearly from 0.19 to 0.34$m_e$ for electrons and from 0.24 to 0.47$m_e$ for holes as the increase of Cl content, indicating a reduction of carrier mobilities upon the Cl addition to \ce{MAPbI3}. The linear fitting produces the formulas $m^*_e(x)=(0.199+0.140x)m_e$ for electrons and $m^*_h(x)=(0.231+0.203x)m_e$ for holes, respectively.

\subsection{Optical properties}\label{opto}
\begin{figure}[!th]
\begin{center}
\includegraphics[clip=true,scale=0.47]{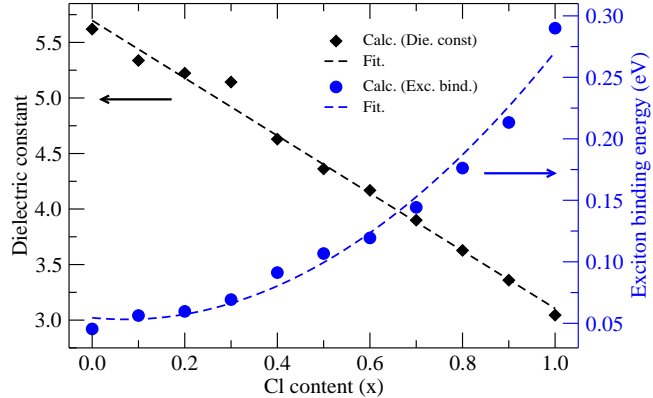}
\end{center}
\caption{\label{fig_excit}Static dielectric constants and exciton binding energies as functions of Cl content x in \ce{MAPb(I_{1-x}Cl_x)3}.}
\end{figure}

In order to estimate the optical properties such as the exciton binding energy and photo-absorption coefficient, we first determined the macroscopic dielectric constants as functions of incident photon energy by solving the Bethe-Salpeter equation (BSE) including the electron-hole interaction ({\it i.e.,} excitonic effect) with or without local field effect (LFE) (see Fig.S5(a) and (b)\dag)~\cite{Onida}. Then the static dielectric constants are extracted from those at the zero photon energy, and they, together with the effective masses of carriers as discussed above, are utilized to determine the exciton binding energy using Eq.~\ref{eig_exciton}. The obtained static dielectric constants and exciton binding energies as functions of Cl content in \ce{MAPbX3} are shown in Fig.~\ref{fig_excit}.

For \ce{MAPbI3}, the static dielectric constant 5.6 in this work is in good agreement with the recent theoretical values of $\sim$4 calculated by quantum molecular dynamics (QMD) method~\cite{Hakamata} and 5.4 by PBE calculation within density functional perturbation theory (DFPT)~\cite{Jong}, and the experimental values of 7-10 measured at 10$^{12}$ Hz ($\sim$0.01 eV)~\cite{Stranks15}. However, larger values were also derived from theoretical calculations, such as $\sim$15 by nonadibatic QMD~\cite{Hakamata} and 18-24 by DFPT with PBEsol functional~\cite{Brivio}. It should be noted that the Bethe-Salpeter approach considering the excitonic and many-body effects in this work gives the almost identical dielectric constants with those from DFPT, which includes the effect of atomic displacements. Using the calculated static dielectric constant and effective masses of electron and hole for \ce{MAPbI3}, the exciton binding energy is calculated to be 45 meV, with an effective exciton Bohr radius of 2.8 nm, which is comparable with the experimental values of 50 meV~\cite{Innocenzo} and 16-45 meV~\cite{Miyata} (see ref.~\cite{Sum} for review). These are low enough to be comparable with the inorganic thin-film semiconductors ($<50$ meV). Nonetheless, taking larger dielectric constants from other theoretical works, the exciton binding energy can drop to 12 meV~\cite{Hakamata} or 10 meV~\cite{Stranks15}, evidencing the existence of free carriers in \ce{MAPbI3}.

It was found in this work that, as the increase of Cl content in \ce{MAPbX3}, the static dielectric constants decrease linearly with a function of $\varepsilon_s(x)=5.698-2.593x$, and accordingly, the exciton binding energies increase quadratically with a function of $E^\text{ex}_b(x)=0.055-0.036x+0.253x^2$ (eV). In the explanation of these variation tendencies, the role of MA cation should be emphasized. In fact, the low frequency regime of the macroscopic dielectric functions can be dominated by rotational motion of MA molecule that has intramolecular dipole, while the high frequency behaviour is related to vibrational polar phonons of the lattice~\cite{Even14}. Here, the MA molecular dipoles can screen the Coulombic interaction between photo-excited electrons and holes in the Pb and I sublattices~\cite{Hakamata}, leading to a reduction of exciton binding. Considering that MA cation is placed in a cuboctahedral cage formed by the nearest 12 halogen atoms, the size of this cage reduces when I atom is replaced by Cl atom ($-25$\%). This induces a restriction of molecular motion, resulting in a decrease of static dielectric constant (reduction of exciton screening) and inversely a sizable increase of exciton binding energy. Here, it is the most interesting that the minimum of exciton binding energy is found at x$\approx$0.07 unlike in the case of \ce{MAPb(I_{1-x}Br_x)3} where it increases linearly~\cite{Jong}. This indicates an enhancement of the light harvesting efficiency when adding a small amount of Cl atom ($\sim$7\%) to \ce{MAPbI3}.

\begin{figure}[!th]
\begin{center}
\includegraphics[clip=true,scale=0.24]{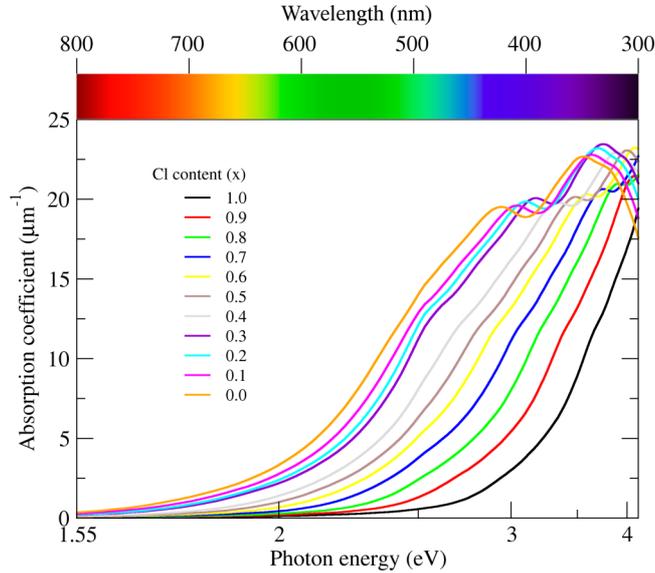}
\end{center}
\caption{\label{fig_absorption}Photo-absorption coefficients of mixed halide perovskites \ce{MAPb(I_{1-x}Cl_x)3} with different Cl content x.}
\end{figure}
Next, using the macroscopic dielectric functions, we obtained the photo-absorption coefficients of mixed iodide-chloride perovskites \ce{MAPbX3}, direct estimation of light harvesting capability of PSC, as shown in Fig.~\ref{fig_absorption} with relevant photon energy range and in Fig.S5(c)\dag~with full photon energy range. It should be noted that the absorption lines in Fig.~\ref{fig_absorption} are from BSE with no LFE for the sake of clear view of variation tendency, while those with LFE for \ce{MAPbI3} and \ce{MAPbCl3} are shown in Fig.S5(c)\dag~for comparison. As can be expected from the band gaps discussed above, when increasing the Cl content in \ce{MAPbX3}, the absorption onset and first peak shift to a higher photon energy, {\it i.e.,} a shorter wavelength light, implying a lowering of power conversion efficiency.

\subsection{Material stability}\label{stabil}
We have already discussed the material stability of mixed iodide-chloride perovskites \ce{MAPb(I_{1-x}Cl_x)3} in subsection~\ref{struct} using lattice related properties such as lattice constant, Pb-X bond length, bulk modulus and cohesive energy. However, it might be thought that these data can not give a direct estimation of material stability. Therefore, we now study their stability by calculating the formation enthalpy $\Delta H$ with respect to the following chemical reaction:
\begin{equation}
\label{form_react}
\ce{MAPbX3} \leftrightarrow \ce{MAX}+\ce{PbX2}
\end{equation}
Here, the formation enthalpy can be calculated as follows:
\begin{equation}
\label{eq_form}
\Delta H_f=H(\ce{MAPbX3})-[H(\ce{MAX})+H(\ce{PbX2})]
\end{equation}
Considering that $H=E+PV$ and the $PV$ term for solid at the atmospheric pressure is in order of $\sim$0.1 meV, we can write $\Delta H\approx\Delta E$ in safe. It is worth noting that the accuracy of the formation enthalpy could be improved to be 0.1 meV per cell due to a cancellation of errors. Although different chemical reactions are thought to be possible such as further reaction of $\ce{CH3NH3PbX3}\leftrightarrow\ce{CH3NH2}+\ce{HX}+\ce{PbX2}$~\cite{Zhang,NiuRev} or even $\ce{CH3NH3PbX3}\leftrightarrow\ce{NH4X}+\ce{H2C}+\ce{PbX2}$~\cite{Mellouhi}, we preferred the reaction Eq.~\ref{form_react} based on the majority of experimental procedures for synthesis of OTPs.

\begin{figure}[!th]
\begin{center}
\includegraphics[clip=true,scale=0.23]{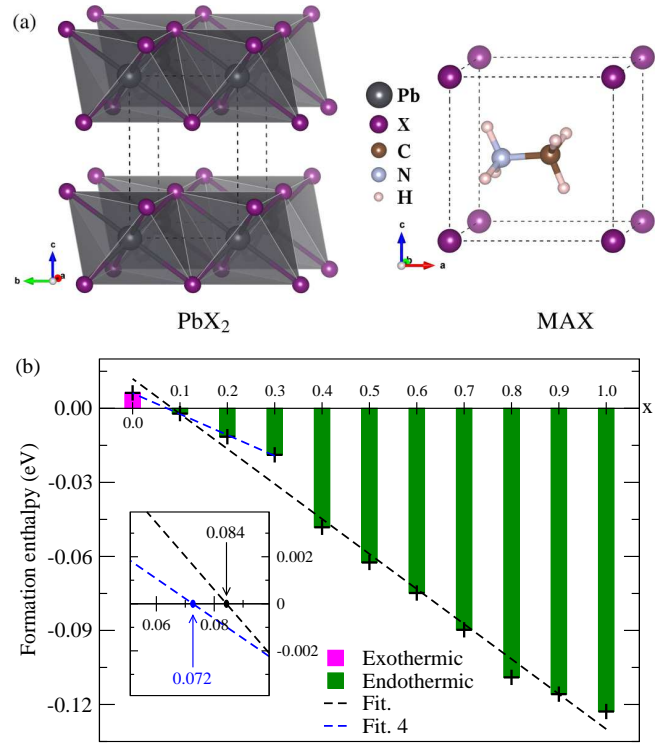}
\includegraphics[clip=true,scale=0.47]{fig5b.eps}
\end{center}
\caption{\label{fig_form}(a) The crystalline structures of rhombohedral \ce{PbX2} and pseudo-cubic \ce{MAX}. (b) Formation enthalpy of \ce{MAPb(I_{1-x}Cl_x)3} from their components \ce{PbX2} and MAX as the increase of Cl content, where the inset shows the intersection points of fitting (dashed) lines with X-axis.}
\end{figure}
We assumed that \ce{PbX2} can be crystallized in rhombohedral structure with space group $P$\={3}$m1$~\cite{Zhang}, which can be characterized by a two-dimensional structure consisting of \ce{PbX2} edge sharing octahedron in a hexagonal arrangement, as shown in Fig.~\ref{fig_form}(a). On the other hand, \ce{MAX} can be crystallized in rocksalt structure, which should be also pseudo-cubic with space group $Pm$ due to the existence of MA cation and its orientational freedom like in \ce{MAPbX3} (Fig.~\ref{fig_form}(a)). Therefore, we performed the structural optimization, maintaining the cubic symmetry and taking account of the MA orientation to confirm the lowest energy configuration. In the context of structural characteristics of both phases, the vdW correction should be considered; the semi-empirical vdW function (vdW-D2) was used. To make it clear that such decomposition of the OTPs causes the deterioration of PSC, we calculated the band gaps of \ce{PbX2} (Fig.S6\dag) and MAX. They are 2.5-3.1 eV in \ce{PbX2} and $\sim$4 eV in MAX, indicating a blue-shift of light absorption onset.

The calculated formation enthalpies for \ce{MAPbX3} as the increase of Cl content are presented with the bar diagram in Fig.~\ref{fig_form}(b). It turns out that only the \ce{MAPbI3} has the positive formation enthalpy, while all other compounds treated in this work have negative values, increasing in the magnitude as almost linear function of Cl content. This result directly indicates that in the former case the phase decomposition reaction is exothermic, while in the latter cases they are endothermic. This means that the phase decomposition of \ce{MAPbI3} into MAI and \ce{PbI2} may occur spontaneously without application of any extrinsic factors such as moisture, oxygen, UV light and elevated temperature, but for other mixed iodide-chloride perovskites the decompositions are thermodynamically unfavorable. It should be noted that in spite of exothermicity of phase decomposition of \ce{MAPbI3} the kinetic barrier can hinder the reaction, explaining its stable existence in a certain period~\cite{Zhang}. When compared with other recent PBE/vdW calculations (0.111/0.119 eV f.u.$^{-1}$ (formular unit) for \ce{MAPbI3} and $-0.040/-0.004$ eV f.u.$^{-1}$ for \ce{MAPbCl3})~\cite{Zhang}, our results (0.006 and $-0.109$ eV, respectively) are in reasonable agreement with them. Moreover, our results are consistent with the experimental findings by Buin {\it et al.}~\cite{Buin15}, reporting that time evolution of the main XRD (X-ray diffraction) peak of \ce{MAPbI3} shows the \ce{PbI2} peak as soon as \ce{MAPbI3} is formed, while \ce{MAPbCl3} does not give any evidence of PbCl$_2$ peak.

We also performed fitting of the calculated formation enthalpies into linear function. This process outputs the linear functions $\Delta H_f(x)=0.0062-0.0846x$ (eV f.u.$^{-1}$) for the first four data from $x=0.0$ to $x=0.3$ (dashed blue line in Fig.~\ref{fig_form}(b)) as well as $\Delta H_f(x)=0.0120-0.1419x$ (eV f.u.$^{-1}$) for the whole range of Cl content (dashed black line). As shown in the inset of Fig.~\ref{fig_form}(b), these lines intersect with X-axis at 0.072 and 0.084, respectively. These are turning points at which the phase decomposition reaction becomes to be endothermic from exothermic. Here we should stress that the turning point $\sim$0.07 is consistent with what we have found for the minimum of exciton binding energy as discussed in subsection~\ref{opto}. Therefore, it can be concluded that at the mixing ratio of Cl $\sim$0.07 into \ce{MAPbI3} the best efficiency could be realized with the positive material stability. We should note that it is much smaller than the one in \ce{MAPb(I_{1-x}Br_x)3} ($\sim$0.2)~\cite{Jong}, and this small ratio value reasonably corresponds to the experimentally observed Cl doping ratio 3-4\%~into \ce{MAPbI3}~\cite{Colella}.

\begin{figure*}[!th]
\begin{center}
\includegraphics[clip=true,scale=0.3]{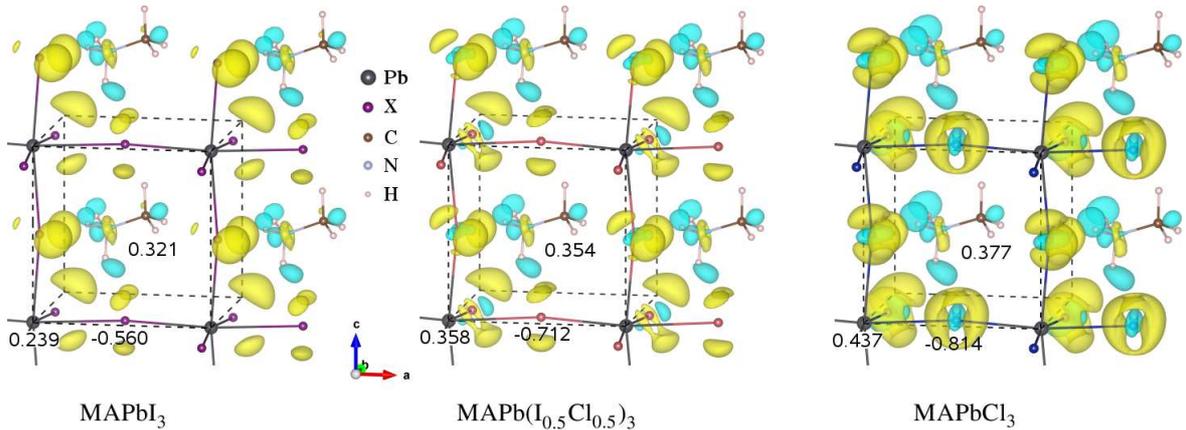}
\end{center}
\caption{\label{fig_den}Electronic charge density difference in \ce{MAPbX3} at the Cl contents of $x=0.0$, 0.5, 1.0. The 3D isosurfaces are plotted at the value of $\pm$0.0017 $|e|$ \AA$^{-3}$, where the yellow (cyan) colour represents the positive (negative) value indicating electron accumulation (depletion). Numbers denote the transferred electronic charges for Pb, X$_3$ and MA molecule calculated by using the Hirshfeld method.}
\end{figure*}
The material stability can be related with the strength of chemical bonds between the constituents of the compound. As discussed in subsection~\ref{struct}, the strength of chemical bonds increase as the increase of Cl content due to the contraction of lattice constant and Pb-X bond length. Besides, since the strength of chemical bond can be inspected by electronic charge transfer in the formation of compound from its components, we estimated the electronic charge density difference between the \ce{MAPbX3} and \ce{PbX3} framework plus MA molecule as follow:
\begin{equation}
\label{chargdf}
\Delta n(\mathbi{r})=n_\ce{MAPbX3}(\mathbi{r})-[n_\ce{PbX3}(\mathbi{r})+n_\ce{MA}(\mathbi{r})]
\end{equation}
For the sake of clear comparison, the electronic charge density differences only at the Cl contents of $x=0.0$, 0.5 and 1.0 are shown in Fig.~\ref{fig_den}. It is observed that a charge accumulation occurs in the interspatial region surrounding the halogen atoms, while a charge depletion occurs around the hydrogen atoms of MA cation and the lead atoms, being enhanced from iodide to chloride. This indicates that electrons are in general transferred from MA cation and Pb atom to halogen atoms, making the chemical bond strong, and the chemical bond gets stronger going from iodide to chloride. To quantitatively estimate the charge transfer, we calculated the atomic charges of atoms by using the Hirshfeld method. As shown in Fig.~\ref{fig_den} and Table S1\dag, going from $x=0.0$ to $x=0.5$ and to $x=1.0$, Pb atom loses electron of 0.239, 0.358 and 0.437, MA molecule also loses electron of 0.321, 0.354 and 0.377, while halogen atoms gain electron of $-0.560$, $-0.712$ and $-0.814$, again indicating the enhancement of charge transfer when mixing chloride with \ce{MAPbI3}. It should be noted that carbon and nitrogen atoms inside MA molecule also gain electrons.

\section{Conclusions}\label{concl}
With {\it ab initio} calculations using the vdW-D2 functional and the efficient VCA method, the effects of Cl addition on structural, electronic, optical properties and material stability of mixed iodide-chloride perovskites \ce{MAPb(I_{1-x}Cl_x)3} were systematically investigated. As the increase of Cl content, the pseudo-cubic crystalline lattice contracts, seeing the decrease of lattice constants according to the linear function of $a(x)=6.310-0.716x$ (\AA) and the increase of bulk modulus with the linear function of $B(x)=10.627+5.492x$ (GPa), which indicates the enhancement of chemical stability. The electronic structure calculations show the increase of band gaps with the quadratic function of $E_g(x)=1.521+0.269x+0.873x^2$ (eV) and the almost linear increase of effective masses of carriers. By solving the Bethe-Salpeter equation with local field effect and using the Mott-Wannier model, we have found that the static dielectric constants decrease linearly according to the function of $\varepsilon_s(x)=5.698-2.593x$, while the exciton binding energies increase quadratically according to the function of $E^\text{ex}_b(x)=0.055-0.036x+0.253x^2$ (eV) with the minimum of 0.054 eV at $x\approx 0.071$. Furthermore, this Cl content $\sim$7\%~is consistent with the turning point at which the decomposition reaction changes to be from exothermic to endothermic, indicating that the best performance can be realized at the Cl addition of $\sim$7\%~to \ce{MAPbI3}. This work provides the evidence why a little addition of Cl to \ce{MAPbI3} enhances the stability as well as the efficiency of PSC and thus reveals new prospects for designing of stable, high efficiency PSCs.

\section*{Acknowledgments}
This work was supported partially by the State Committee of Science and Technology, DPR Korea, under the state project `Design of Innovative Functional Materials for Energy and Environmental Application' (no.2016-20). The calculations have been carried out on the HP Blade System C7000 (HP BL460c) that is owned and managed by the Faculty of Materials Science, Kim Il Sung University.

\section*{Appendix A. Supplementary data}
Supplementary data related to this article can be found at URL.

\section*{\label{note}Notes}
The authors declare no competing financial interest.

\bibliographystyle{elsarticle-num-names}
\bibliography{Reference}

\end{document}

\title{Supplementary information -- Revealing the stability and efficiency enhancement in mixed halide perovskites \ce{MAPb(I_{1-x}Cl_x)3} with {\it ab initio} calculations}

\author{Un-Gi Jong,$^{ab}$ Chol-Jun Yu,$^a$\footnote{Corresponding author: Chol-Jun Yu, ryongnam14@yahoo.com} Yong-Man Jang,$^b$ Gum-Chol Ri,$^a$ Song-Nam Hong,$^a$ Yong-Hyon Pae$^c$ \\
\small \it $^a$ Department of Computational Materials Design, Faculty of Materials Science, Kim Il Sung University, \\ 
\small \it Ryongnam-Dong, Taesong District, Pyongyang, Democratic People's Republic of Korea, \\
\small \it $^b$ Natural Science Centre, Kim Il Sung University, Ryongnam-Dong, \\
\small \it Taesong District, Pyongyang, Democratic People's Republic of Korea\\
\small \it $^c$ Institute of Physics Engineering, Kim Chaek University of Technology, \\
\small \it Central District, Pyongyang, Democratic People's Republic of Korea }
\date{}
\maketitle

\begin{figure}[!th]
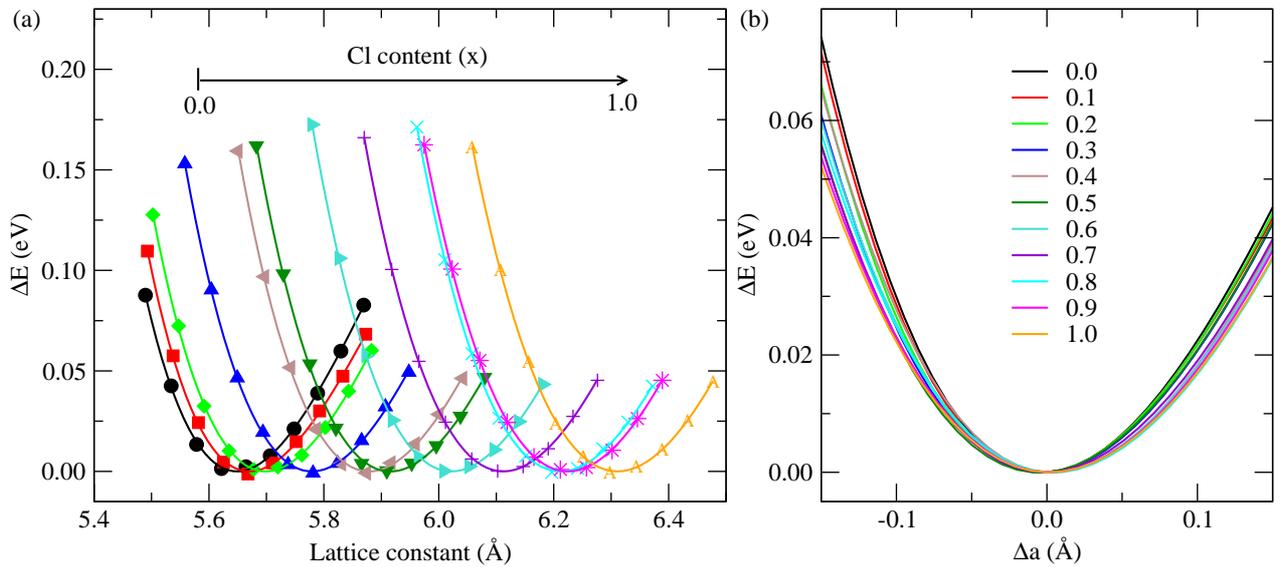

\begin{center}
\includegraphics[clip=true,scale=0.55]{figs1a.eps}
\includegraphics[clip=true,scale=0.55]{figs1b.eps}
\end{center}
\caption{\label{figs1}$E-a$ curves obtained by fitting into the Birch-Murnaghan equation of state for \ce{MAPb(I_{1-x}Cl_x)3} (x=[0, 1]). (a) The energy of equilibrium state is set to be zero in y-axis, and (b) the equilibrium lattice constant is also set to be zero in x-axis.}
\end{figure}
\pagebreak

\begin{figure}[!th]
\begin{center}
\begin{tabular}{lc}
\includegraphics[clip=true,scale=0.5]{figs2a.eps} &
\includegraphics[clip=true,scale=0.5]{figs2b.eps}\\
\includegraphics[clip=true,scale=0.5]{figs2c.eps} & \includegraphics[clip=true,scale=0.15]{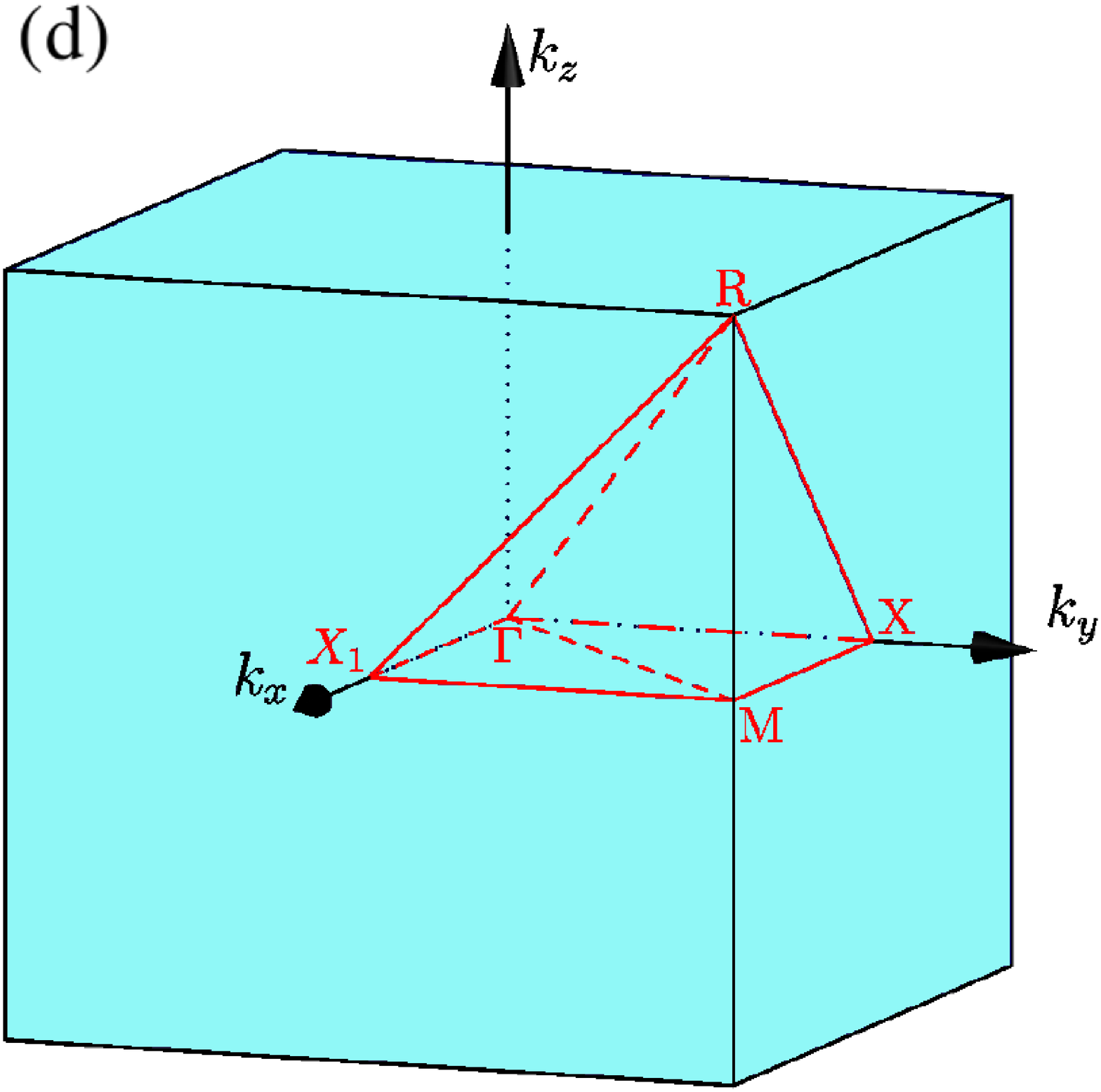}\\
\end{tabular}
\end{center}
\caption{\label{figs2}Electronic band structures of \ce{MAPb(I_{1-x}Cl_x)3} according to high symmetry points and lines in the first Brillouin zone (d). (a) x=0.0, (b) x=0.5, (c) x=1.0. }
\end{figure}
\pagebreak

\begin{figure}[!th]
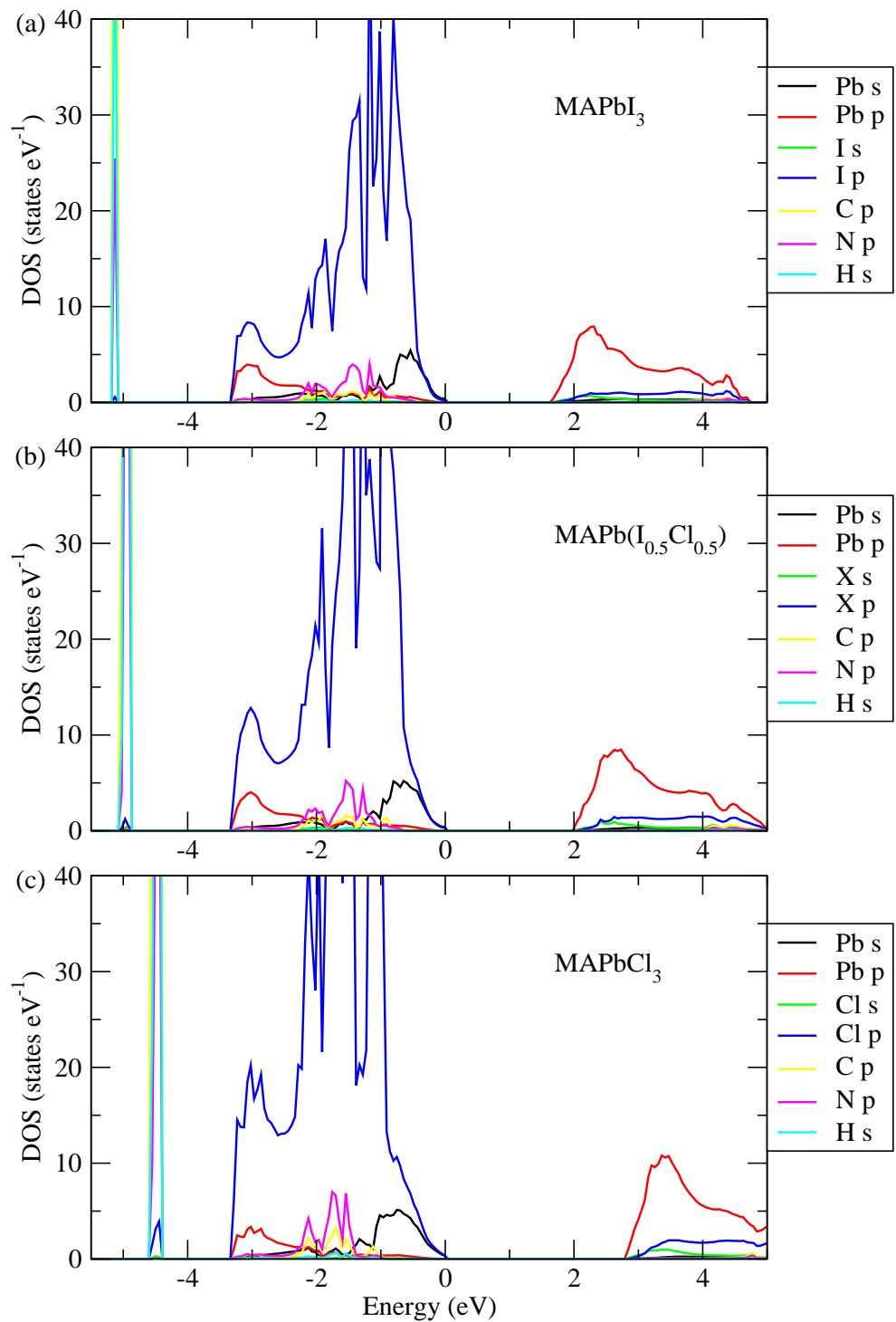

\begin{center}
\includegraphics[clip=true,scale=0.65]{figs3a.eps}
\includegraphics[clip=true,scale=0.65]{figs3b.eps}
\includegraphics[clip=true,scale=0.65]{figs3c.eps}
\end{center}
\caption{\label{figs3}Atmoic resolved and partial density of states in \ce{MAPb(I_{1-x}Cl_x)3}. (a) x=0.0, (b) x=0.5 and (c) x=1.0.}
\end{figure}
\pagebreak

\begin{figure}[!th]
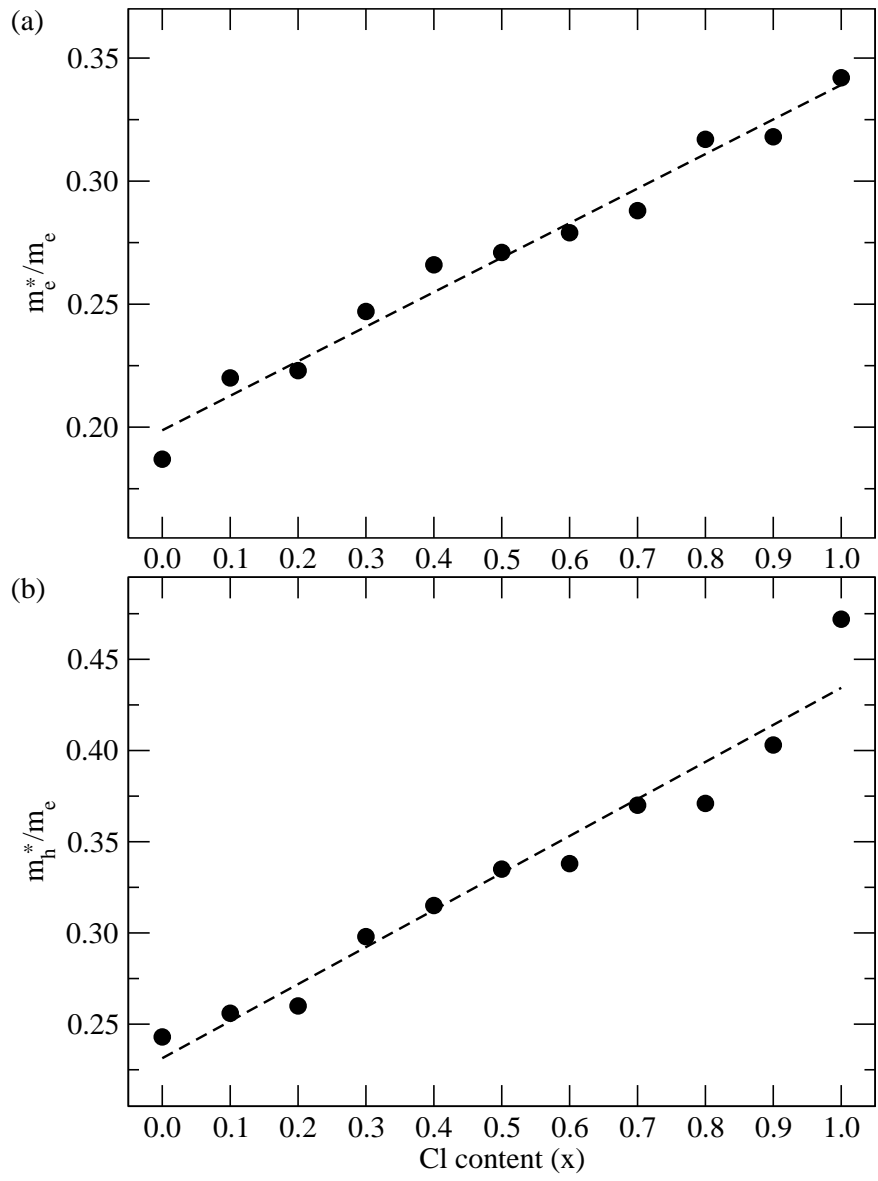

\begin{center}
\includegraphics[clip=true,scale=0.65]{figs4a.eps}
\includegraphics[clip=true,scale=0.65]{figs4b.eps}
\end{center}
\caption{\label{figs4}Effective masses of electron and hole for \ce{MAPb(I_{1-x}Cl_x)3}. (a) $m_e^*/m_e$, (b) $m_h^*/m_e$.}
\end{figure}
\pagebreak

\begin{figure}[!th]
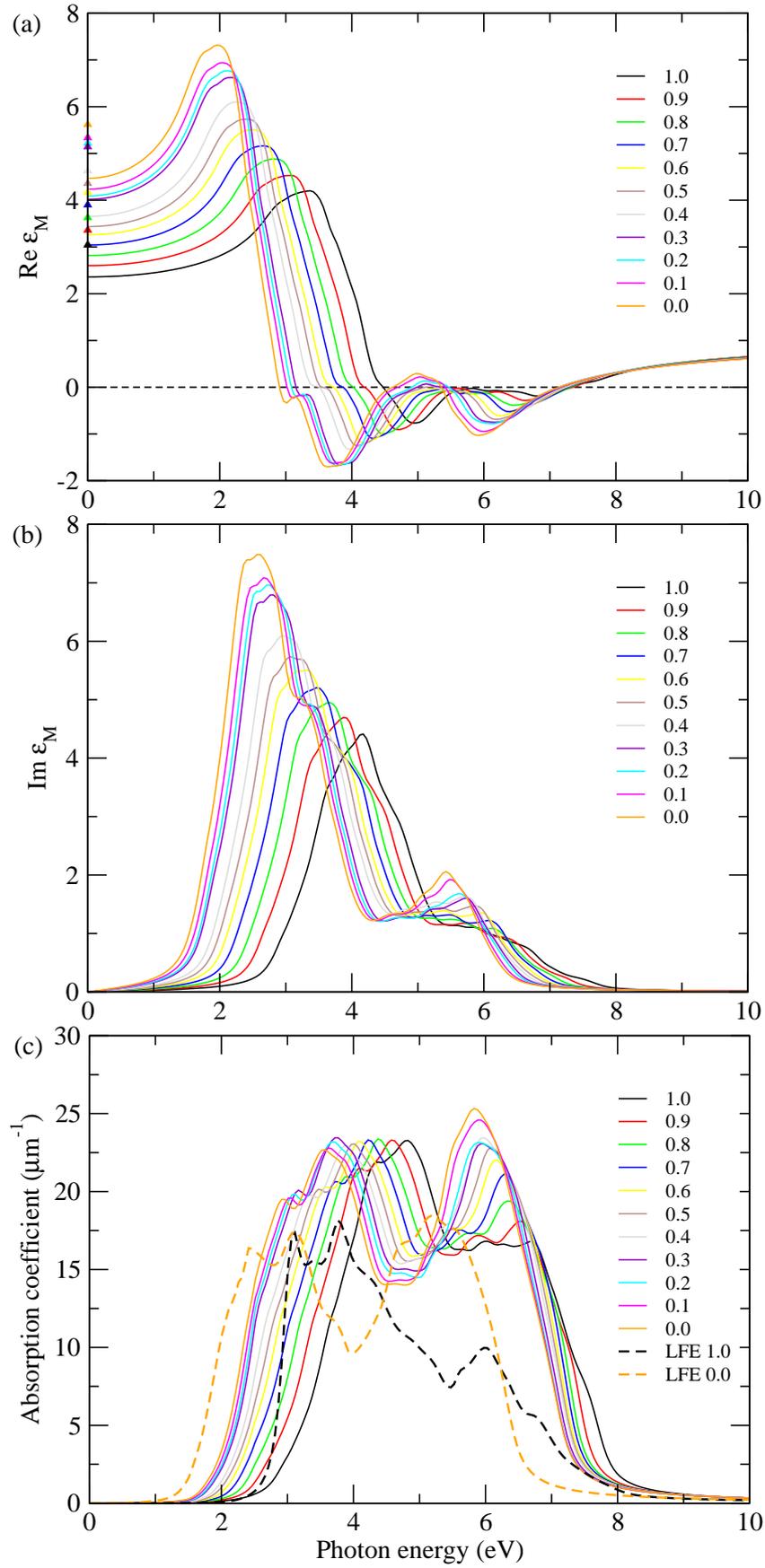

\begin{center}
\includegraphics[clip=true,scale=0.64]{figs5a.eps}
\includegraphics[clip=true,scale=0.64]{figs5b.eps}
\includegraphics[clip=true,scale=0.64]{figs5c.eps}
\end{center}
\caption{\label{figs5} (a) Real part and (b) imaginary part of macroscopic dielectric functions, and (c) photo-absorption coefficients, as the increase of Cl content, obtained by the Bethe-Salpeter approach with no local field effect (LFE). Triangles in (a) and dashed lines in (c) are those with LFE with scissor operator energies of 0.0 for \ce{MAPbI3} and 0.6 eV for \ce{MAPbCl3}.}
\end{figure}
\pagebreak

\begin{figure}[!th]
\begin{center}
\includegraphics[clip=true,scale=0.64]{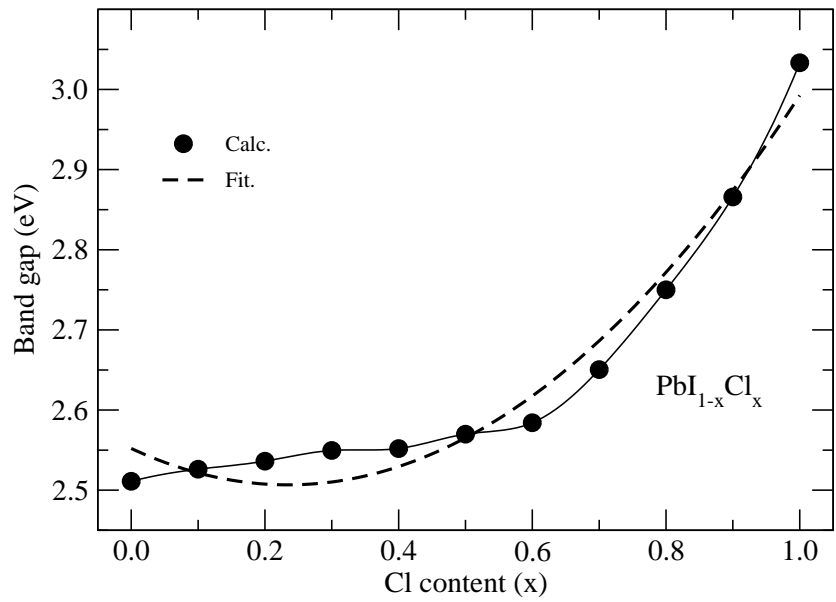}
\end{center}
\caption{\label{figs6}Band gaps for \ce{PbX2} with rhombohedral structure.}
\end{figure}
\begin{table}[!th]
%\centering
\caption{\label{tab_hir}Atomic charge of atoms in \ce{MAPb(I_{1-x}Cl_x)3} (x=0.0, 0.5, 1.0), calculated by using the Hirshfeld method. The sum of atomic charges of some relevant atoms is also presented. Positive value indicates the loss of electron, while negative value means the gain of electron. }
\begin{tabular}{lrrr}
\hline
Atom &  0.0~~ & 0.5~~ & 1.0~~ \\
\hline
Pb &  0.239 &  0.358 &  0.437\\
X  & -0.151 & -0.201 & -0.237 \\
   & -0.230 & -0.283 & -0.314 \\
   & -0.179 & -0.229 & -0.263 \\
(X$_3$) & -0.560 & -0.712 & -0.814 \\
C  & -0.078 & -0.075 & -0.074 \\
N  & -0.045 & -0.041 & -0.037 \\
H  &  0.105 &  0.112 &  0.118 \\
   &  0.105 &  0.112 &  0.118 \\
   &  0.111 &  0.116 &  0.118 \\
   &  0.042 &  0.044 &  0.044 \\
   &  0.042 &  0.044 &  0.044 \\
   &  0.039 &  0.042 &  0.045 \\
(MA)&  0.321 &  0.354 &  0.377 \\
\hline
\end{tabular} \\
\normalsize
\end{table}